\documentstyle[epsfig]{l-aa}
\begin{document}
   \thesaurus{11                   
              (11.17.1;            
               11.17.4 1556+3517;  
               13.09.1)}           
   \title{ISO observations of the radio-loud BALQSO 1556+3517 \thanks
    {Based on observations with ISO, an ESA project with Instruments funded
    by ESA member states (especially the PI countries: France, Germany, The
    Netherlands and the United Kingdom) with the participation of ISAS and 
    NASA}}


   \author{J. Clavel
          }

   \offprints{J. Clavel (jclavel@iso.vilspa.esa.es)}

   \institute{ESA ISO Observatory, P.O. Box 50727, 28080 Madrid, Spain
             }

   \date{Received: 12 September 1997; accepted: 25 Novemnber 1997}

   \maketitle

   \begin{abstract}

   To date, the quasar 1556+3517 is the only radio loud Broad Absorption 
   Line QSO (BALQSO; Becker et al. 1997). This prompted narrow band 
   filter imaging observations in the range 4--15~$\mu$ with the ISO 
   satellite. The source is clearly detected in all filters and appears 
   point-like and isolated at the resolution of ISO. The overall Spectral 
   Energy Distribution (SED) in ${\rm erg~s^{-1}}$ peaks at a rest wavelength 
   $\geq~6~\mu$ and is reddened by 1.6 visual magnitudes. Correction for 
   reddening brings 1556+3517 within 1.3 sigma from the ${\rm \log R\,=\,1}$ 
   dividing line between radio-loud and radio-quiet sources. The 
   mid-IR luminosity integrated over the 1.8--6~$\mu$ 
   rest wavelength range is ${\rm \sim\,2\,10^{46}\,erg\,s^{-1}}$ 
   (${\rm H_{0}\,=\,75\,km\,s^{-1}\,Mpc^{-1},~q_{0}\,=\,0}$), which requires 
   at least ${\rm 4\,M_{\sun}}$ of dust and ${\rm 800\,M_{\sun}}$ of
   associated gas. It is unlikely that such a large mass stems from the
   BAL wind itself. 

   \end{abstract}

%
%

\section{Introduction}

Broad Absorption Lines (BAL) quasars constitute about $\sim$ 10~\% of the
overall quasar population. Observationally, they are characterized by very
broad ($0.1~c$ at zero intensity) and blue shifted absorption lines mostly 
from highly ionized species such as NV$\lambda$1240, CIV$\lambda$1550, \ldots. 
Approximately 10~\% of the BALQSO display in addition absorption lines from 
low ionisation species such as SiII, MgII and AlIII (Weymann et al. 1991).
These are the so-called low ionisation BALQSO, or Lo-BAL. BALQSO in 
general are also heavily absorbed in the X-rays (Green \& Mathur 1996) though 
it is not clear that the hot and highly ionized gas needed to produce the 
ionization edges seen in the keV range is the same as that which produces 
the UV and optical absorption troughs.

The BALQSO phenomenon is poorly understood theoretically. In particular, its
relation to the general QSO phenomenon is unclear. There are basically two 
classes of models: in the first class, BALQSO represent a transitory 
``cocoon'' phase in an evolutionary process whereby a temporary excess of 
accretion fuel is expelled by radiation pressure from the central engine 
through a wind (Voit, Weymann \& Korista 1993). This model predicts that 
eventually BALQSOs will evolve into ``normal'' quasars once the excess gas 
has been removed. In the second class of models, BALQSO are normal quasars 
but viewed at a particular inclination such that our line-of-sight intercept 
a wind of material blown-off the surface of the accretion disk or the top of 
the molecular torus (e.g. Murray et al. 1995).
Spectropolarimetric observations tend to support this second hypothesis 
in that they often reveal significantly polarized emission which is best 
explained by the scattering of a partially obscured but otherwise normal 
quasar spectrum (e.g. Wills et al. 1992, Cohen et al. 1995, Hines \& Wills 
1995, Goodrich \& Miller 1995 \ldots). 
The fact that the emission lines and continuum properties of BALQSOs are 
indistinguishable from that of non BAL QSOs (Weymann et al. 1991) also 
lend credit to the viewing direction hypothesis.

An intriguing and important property of BALQSOs is that, up to now, there were
no known radio-loud broad absorption line quasars (Stocke et al. 1992). The 
criterion commonly used for radio-loudness is ${\rm \log R \geq 1}$, where R
is the K--corrected ratio of 5~GHz to 4400\,\AA\/ (B Band) flux (Stocke et al. 
1992). There is some intriguing suggestion that BALQSOs show an excess of 
radio emission over non-BALQSOs such that there is an apparent accumulation of
BALQSOs close-to but slightly below the ${\rm \log R~=~1}$ dividing line
(Francis, Hooper and Impey 1993). 

The quasar 1556+3517 (z = 1.48) was recently discovered in the course of 
the FIRST VLA radio survey (Becker White and Helfand 1995). 1556+3517
belongs to the Lo-BAL class, and, with $\log R~=~3.1$, unambiguously qualifies 
as the first and so far only radio-loud BALQSO (Becker et al. 1997). 
Because it holds the promise of shedding light on the radio-loudness/BALQSO 
dichotomy and, by extension, on the difference between radio--loud and 
radio--quiet quasars, I initiated a program to observe 1556+3517 in the 
mid-IR with the ISO satellite (Kessler et al. 1996). One particular aim was 
to obtain a better estimate of the amount of dust on the line of sight to the 
QSO and check that 1556+3517 remains radio-loud even after reddening
correction. Another was to verify the prediction that Lo-BAL QSO in general
are heavily obscured. More generally, it is important to measure the spectral 
energy distribution (SED) of BALQSO over as wide a frequency range as 
possible in order to compare it to that of non-BAL QSO in the hope of finding
differences that hold clues to the origin of the BAL phenomenon.


\section{Observations and data reduction}

   The quasar 1556+3517 was observed with the $32\times 32$~px long wavelength 
   (LW) array of the ISOCAM instrument (Cesarsky et al. 1996) on board the 
   ISO satellite (Kessler et al. 1996). The observation was executed 
   nominally with the C01 observing mode (AOT) on 14 March 1997 (revolution 
   484), starting at 06:17:10 (UT) and lasting 46 minutes and 26 second. 
   It consisted of a series of $3\times 3$ raster maps centered on the 
   optical position of the QSO (R.A. = 15h~56m~33.8s, Dec = 
   35$\degr$~17$\arcmin$~58$\arcsec$, epoch = J2000), with one map per 
   filter. The magnification selected was 3$\arcsec$ per pixel and the 
   distance between two raster points was 6$\arcsec$. The axis of the 
   rasters were oriented parallel to the celestial North and East. The number 
   of exposures per raster dwell point was 19, with a unit integration time of 
   2.1 s per exposure and an amplifier gain of 2. The 6 filters used were, in 
   sequential order: LW3 [12--18~$\mu$], LW2 [5--8.5~$\mu$], LW9 
   [14--16~$\mu$], LW8 [10.7--12~$\mu$], LW7 [8.5--10.7~$\mu$] and LW1 
   [4--5~$\mu$], where the numbers in brackets refer to the wavelength
   range covered by each of the filter. This sequence corresponds
   to decreasing pixel illuminations such that it minimizes the detector
   stabilization time. 
   
\begin{figure}[t]
   \begin{center}
   \centerline{\epsfig{file=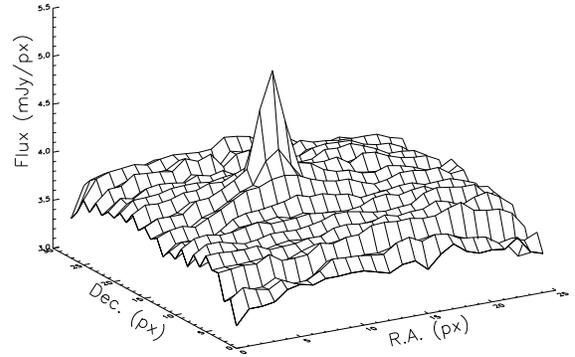,width=8.0cm}}
   \end{center}
      \caption[]{The 12--18~$\mu$ map of the field
              around the quasar 1556+3517; the size of the map is 
              $75\times 84~\arcsec$ and the scale ${\rm 3~\arcsec\,px^{-1}}$.}
         \label{map}
   \end{figure}

   The six monochromatic maps were processed with the CAM Interactive 
   Analysis (CIA) Software \footnote{CIA is a joint development by the ESA 
   Astrophysics Division and the ISOCAM Consortium led by the ISOCAM PI, 
   C. Cesarsky, Direction des Sciences de la Matiere, C.E.A., France.} 
   starting from the Edited Raw Data (ERD). The first few ($\sim$ 4) 
   exposures of each frame were discarded so as to retain only those data
   for which the detector throughput had reached 90~\% of its stabilisation 
   level. The remainder of the processing was standard and involved the
   usual steps, i.e. dark current subtraction, flat-fielding, removal of
   particle hits (``de-glitching'') and flux calibration. The source is
   clearly detected in all six maps at a very high significance level
   (e.g. 63 sigma for the LW3 map). The 12--18~$\mu$ map is shown in 
   Fig.\ref{map} as an example. 

%
%

   The monochromatic intensities were computed by integration of the flux in 
   a 6~px radius (18$\arcsec$) circle centered on the source and subtraction
   of the normalized background integrated in an annulus of 8 and 13 px 
   internal and external external radii, respectively. The results are
   displayed in table~\ref{results}. The errors quoted derive mainly from the 
   uncertainty in the calibration and background, the latter being induced by 
   residual flat-fielding errors. The full widths at half maximum (FWHM) of the
   source images were computed by fitting a two dimensional gaussian to the flux spatial
   distribution. The results are listed in column 5 of table~\ref{results}. 
   For comparison, the theoretical value of the full width at 41~\%
   of the maximum intensity of a diffraction limited telescope of diameter D is
   ${\rm \frac{\lambda}{D}}$, i.e. 0.343\,${\rm \lambda\,(\mu)}$ in the case of
   ISO. This yields e.g. 5.0~$\arcsec$ for the
   LW3 image, consistent with the value in table~\ref{results}. As a sanity
   check, the LW3 images of point sources obtained with the same pixel
   magnification were  retrieved from the set of ISO calibration files and 
   analysed in the same fashion, yielding FHWM\,=\'4.4~$\arcsec$, in good 
   agreement with the FWHM of 1556+3517. For all six filters, the source width 
   is consistent with that of a point source. Neither of the six maps show
   extended structures nor close-by objects at least down to the detection 
   limits quoted in column 6 of table~\ref{results}. These limits were 
   obtained by computing the standard deviation of the background in areas of 
   the map well outside the image of the source.
   
   \begin{table}
      \caption[ ]{Monochromatic mid IR fluxes of 1556+3517}
         \label{results}
   \begin{flushleft}
   \begin{tabular}{cccccc}
            \hline
            \noalign{\smallskip}
LW & $\Delta \lambda$ & $\bar{\lambda_{rest}}$ & Flux & FWHM & Limit \\
     & ($\mu$)        & ($\mu$)                &(mJy) & ($\arcsec$) & 
       (${\rm \mu\,Jy/\sq}$) \\
            \noalign{\smallskip}
            \hline
            \noalign{\smallskip}
 1   & 4.00--5.00 & 1.81 & $1.5\pm0.4$ & $3.5\pm1.0$ & 14 \\
 2   & 5.00--8.50 & 2.75 & $4.7\pm0.5$ & $3.0\pm0.4$ & 8  \\
 7   & 8.50--10.7 & 3.90 & $5.7\pm0.3$ & $2.9\pm0.5$ & 11 \\
 8   & 10.7--12.0 & 4.56 & $9.8\pm0.5$ & $3.2\pm0.6$ & 20 \\
 3   & 12.0--18.0 & 5.85 & $12.7\pm0.5$& $4.3\pm0.5$ &  8 \\
 9   & 14.0--16.0 & 6.01 & $8.3\pm1.5$ & $3.9\pm1.0$ & 18 \\
            \noalign{\smallskip} 
            \hline
   \end{tabular}
   \end{flushleft}
   \end{table}
%

\section{Discussion and conclusions}

\begin{figure}[t]
   \begin{center}
   \centerline{\epsfig{file=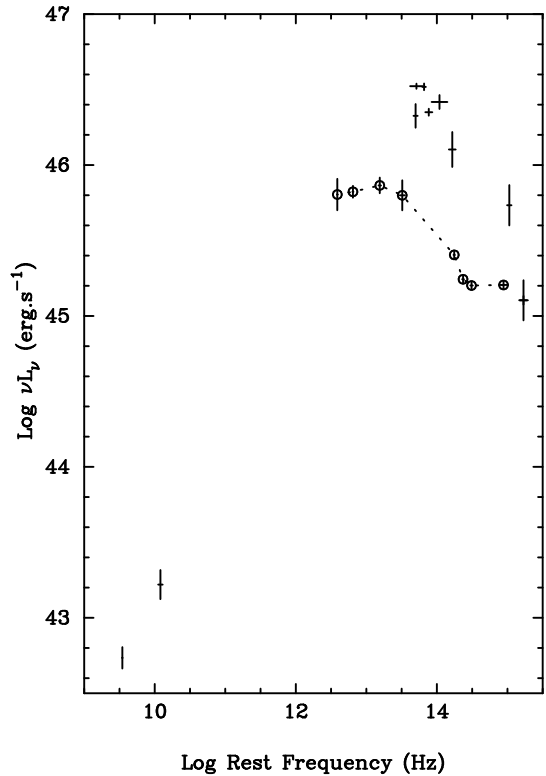, width=8.0cm}}
   \end{center}
      \caption[]{The overall spectral energy distribution (SED) of 1556+3517
                  in the QSO rest frame from radio to ultraviolet.
                  The SED of the IR selected BALQSO IRAS~14026+4341 is also
                  plotted (circled symbols connected by a dotted line)
                  for comparison.}
         \label{SED}
   \end{figure}

 The overall rest frame Spectral Energy Distribution (SED) of 1556+3517,
 from 1.5 GHz to the B band, is plotted in fig.\ref{SED}, where the optical 
 and radio data are taken from Becker et al. (1997). All observed fluxes have 
 been converted to rest frame luminosities $\nu\,L_{\nu\,e}$ (in 
 erg\,s$^{-1}$) assuming  $\rm{H_{o} = 75\,km\,s^{-1}\,Mpc^{-1}}$,  
 $\rm{q_{o} = 0}$ and using the standard equations (e.g. Weedman 1986):

   \begin{equation}
      L_{\nu\,e} = 4\,\pi\,d^2\,(1 + z)\,f_{\nu\,o}
   \end{equation}

   \begin{equation}
      d = \frac{c\,z\,(1 + z/2)}{H_{o}\,(1 + z)}
   \end{equation}

   where L$_{\nu\,e}$ is the monochromatic luminosity per unit frequency 
   in the rest frame of the QSO, f$_{\nu\,o}$ is the monochromatic flux
   per unit frequency in the observer's frame, d is the distance to the QSO
   and the other symbols have their usual meanings. 

%
%

  The SED of the quasar IRAS 14026+4341 is also shown in fig.\ref{SED} for 
  comparison, the data being taken from Low et al. (1989). The comparison is
  illustrative since IRAS~14026+4341 is an IRAS discovered BALQSO with a 
  large IR-to-optical luminosity ratio (Low et al. 1989). It is immediately 
  apparent that, despite similar ultraviolet luminosities, 1556+3517 is almost
  an order of magnitude more luminous in the near to mid-IR than 14026+4341. 

  This presumably indicates the presence of a large amount of warm dust in 
  1556+3517 which causes a depression of the optical-UV continuum by reddening
  and an enhancement of the near and mid-IR flux because of grain thermal
  emission.
  
  The 1556+3517 spectral index (${\rm F_{\nu} \propto \nu^{\alpha}}$) from
  the B band (${\lambda_{rest}\,=\,0.1774~\mu}$) to the LW1 filter 
  (${\lambda_{rest}\,=\,1.81~\mu}$) is $\alpha = -1.99\pm0.25$.
  Using the R band (${\lambda_{rest}\,=\,0.2826~\mu}$) flux instead yields 
  $\alpha = -1.46\pm0.31$. Both indices are significantly steeper than those 
  of ``normal'' non-BAL quasars. Note that only the hottest dust grains at 
  $\sim 1600$~K yield a significant contribution to the 1.81~$\mu$ flux. Since 
  such a temperature is higher or equal to the grain sublimation temperature, 
  relatively little dust thermal emission is expected at this wavelength. 
  Hence, to a first approximation, the steep UV-to-near-IR spectral index of 
  1556+3517 is mostly a consequence of dust extinction.

  There are two reasons why the $\alpha\,=\,-1.99\pm0.25$ index value 
  should be preferred:
  \begin{enumerate}
  \item{The 1774.2~\AA\ band lies in a region relatively free of any strong
   emission or absorption lines. This is not the case of the 2823~\AA\ band
   which is contaminated by the ${\rm MgII\lambda\,2798}$ emission and
   absorption.}
  \item{More importantly, the 2823~\AA\ band is definitely enhanced by
   Balmer continuum (BaC) emission and blended low-contrast FeII emission
   lines which create a pseudo-continuum in quasar spectra often referred to 
   as the ``small--bump''.}
  \end{enumerate}

  The extinction internal to 1556+3517 was therefore computed under the 
  following assumptions:
  \begin{enumerate}
  \item{The ``intrinsic'', i.e. un-reddened, optical-UV spectral index of 
   1556+3517 is -0.5. Within the uncertainty, this is equal to the mean UV 
   spectral index $\alpha\,=-0.46\pm0.05$ of the sample of 33 high redshift 
   (z\,$\geq$\,1) QSOs of O'Brien, Gondhalekar and Wilson (1988) as well as to 
   the mean optical spectral index $\alpha\,=\,-0.5\pm0.1$ of the large 227 
   QSO sample of Cheney and Rowan-Robinson (1981) or that obtained by 
   Neugebauer et al. (1987) from their sample of 104 quasars 
   ($\alpha\,=\,-0.4\pm0.04$).} 
   \item{The reddening law is as parametrized by Seaton (1979) in the 
   optical-ultraviolet range and as tabulated by Rieke and Lebofsky (1985) from
   1 to 13\,$\mu$.}
  \end{enumerate}

  Under these conditions, the amount of visual extinction internal to 1556+3517
  required to steepen the power-law index from -0.5 to $\alpha~=~-1.99$ is
  given by:
  
  \begin{equation}
  A_v\,=\,\frac{(\alpha\,+\,0.5)\,ln(\frac{\lambda_1}{\lambda_2})}
  {ln(10^{0.4\,(R_{\lambda_1}\,-\,R_{\lambda_2}})}
  \end{equation}
  
  where $[\lambda_1-\lambda_2]$ is the (rest) wavelength range over which
  the spectral index is measured (0.177--1.810~$\mu$ in the present case) and
  the function $R(\lambda)\,=\,\frac{A_{\lambda}}{A_V}$ is the reddening law 
  at wavelength $\lambda$. This yields a visual extinction 
  ${\rm A_{v}\,=\,1.64\pm0.27}$ magnitudes which implies that the continuum is 
  reddened by 4.01
  magnitudes at a rest wavelength of 1774~\AA\/ but only 0.256 magnitudes
  at 1.81~$\mu$. Note that the foreground galactic extinction in the direction 
  of 1556+3517 (${\rm l_{II}\,=\,56.48\degr, b_{II}\,=\,49.95\degr}$) is only 
  0.05 V magnitudes and can be neglected.

  After correction for extinction, the radio--to--optical flux ratio becomes
  ${\rm \log R\,=\,1.50\pm0.23}$ if the B band data are used to derive the 
  flux at ${\rm \lambda_{rest}\,=\,4400\,\AA}$ (K correction) and 
  ${\rm \log R\,=\,1.15\pm0.25}$ if one uses the R flux instead. The two 
  estimates differ at the 1.1 sigma level only. I therefore adopt the mean of 
  the two ${\rm \log R\,\simeq\,1.32\pm0.25}$. Such a ratio is less than 1.3 
  sigma away from the ${\rm \log R\,=\,1}$ dividing line between radio-loud 
  and radio-quiet quasars. One therefore concludes that the case for 
  1556+3517 being radio-loud is marginal, at best. It is also worth
  pointing-out that de-reddening brings 1556+3517 close to the 
  ${\rm \log R\,=\,1}$ line where data suggests that there is an apparent
  excess of BALQSOs over non-BALQSOs (Francis, Hooper and Impey 1993).

  The near to mid-IR luminosity ${\rm L_{MIR}}$ of 1556+3517 was computed 
  by integrating all the de-reddened rest-frame emission over the range 
  ${\rm \lambda_{rest}\,=\,1.81\,\mu}$ to ${\rm \lambda_{rest}\,=\,6.01\,\mu}$
  after subtraction of an -0.5 index power-law matching the de-reddened B band
  and LW1 band luminosities: ${\rm L_{MIR}\,=\,1.9\,10^{46}\,erg\,s^{-1}}$.
  
  One can use ${\rm L_{MIR}}$ to estimate the mass of dust responsible for
  near to mid-IR emission in 1556+3517. The mean rest wavelength of the
  MIR emission corresponds to a black-body temperature of $\sim$\,740~K.
  Under the assumptions that the dust grains have a mean size of 0.1~$\mu$ 
  and radiate like black-bodies and following Rudy and Puetter (1982), I 
  obtain a dust mass ${\rm M_{d}\,\sim\,4\,M_{\sun}}$. This should in fact 
  be seen as a lower limit since actual grains are likely to emit less
  efficiently than black-body thereby requiring an even larger mass to 
  produce the observed MIR luminosity. For a normal galactic dust--to--gas 
  mass ratio of 200, this implies a total mass of gas of about 
  ${\rm 800\,M_{\sun}}$. 

  The color temperature of the dust is higher than that of typical HII regions 
  which suggests that it is closely associated to the active nucleus.
  It is therefore legitimate to ask whether the dusty gas responsible for
  the MIR emission could be the BAL wind itself. In the disk wind model of
  Murray et al. (1995), the BAL region covering factor is about 10~\%, its
  column density ${\rm N_{H}\,\sim\,10^{23.5}\,cm^{-2}}$ and its inner 
  radius is of the order of $10^{16}$\,cm. This yields a BAL mass 
  ${\rm M_{BAL}\,\simeq\,3\,10^{-2}\,M_{\sun}}$, more than 4 orders of
  magnitude smaller than the above estimate. Of course, there is more mass
  available as the wind will accumulate material over the years. For an
  axially symmetrical wind, the mass flow rate is given by:
   \begin{equation}
   \dot{M} = 3.33\,10^{-2} (\frac{\Omega}{4 \pi}) N_{23} R_{16} v_{8} 
    \mbox{~~~~(M$_{\sun}$ yr$^{-1}$)}
   \end{equation}
  where, ${\rm \frac{\Omega}{4 \pi}}$ is the solid angle sustained by the 
  wind as seen from the central source,
  ${\rm N_{23}}$ its column density in units of ${\rm 10^{23}\,cm^{-2}}$,
  ${\rm R_{16}}$ its radius in units of ${\rm 10^{16}\,cm}$, and 
  ${\rm v_{8}}$ its velocity in units of ${\rm 10^{8}\,cm\,s^{-1}}$.
  Assuming parameters value as in Murray et al's model yields 
  ${\rm \dot{M}\,\simeq\,3.3\,10^{-3}\,M_{\sun}\,yr^{-1}}$. 
  This implies that the BAL phenomenon should have lasted at least 
  $\sim$~250,000 years to accumulate the MIR emitting gas. While such a 
  value in itself is not unreasonable, there are problems with this scenario:

  \begin{itemize}
   \item{On the one hand, the dust must somehow ``see'' the central UV source 
   in order to retain a temperature of $\sim$ 700~K for such a long time.
   However, given the mass involved and the geometry of the wind, the opacity
   will be extremely high and the bulk of the dust and gas will be 
   (self-)shielded from the central source.}
   \item{This gas would presumably be at least partially ionized. Since the
   total column would far exceed ${\rm 10^{24} cm^{-2}}$, the gas would be
   opaque to free-free absorption and the continuum totally suppressed, 
   contrary to what is observed.}
  \end{itemize}  
  
   It therefore seems unlikely that the observed MIR emission originates
   from the wind itself. A more plausible origin would be e.g., the molecular
   torus, the Narrow Line Region, or a recent burst of star formation.

\begin{acknowledgements}
    It is a pleasure to thank the Time Allocation Committee for awarding 
    discretionary ISO observing time to this project. I am also grateful 
    to the referee, Dr. Ray Weymann, for valuable comments which improved 
    the manuscript.
\end{acknowledgements}

\end{document}